# A First Determination of the Surface Density of Galaxy Clusters at very low X–ray fluxes

Piero Rosati[1,2], Roberto Della Ceca, Richard Burg[2], Colin Norman[2] and Riccardo Giacconi[3,2]

*Department of Physics & Astronomy, The Johns Hopkins University, Baltimore, MD, 21218*

## ABSTRACT

We present the first results of a serendipitous search for clusters of galaxies in deep ROSAT-PSPC pointed observations at high galactic latitude. The survey is being carried out using a Wavelet based Detection Algorithm which is not biased against extended, low surface brightness sources. A new flux–diameter limited sample of 10 cluster candidates has been created from $\sim 3\,\mathrm{deg}^2$ surveyed area. Preliminary CCD observations have revealed that a large fraction of these candidates correspond to a visible enhancement in the galaxy surface density, and several others have been identified from other surveys. We believe these sources to be either low–moderate redshift groups or intermediate to high redshift clusters. We show X-ray and optical images of some of the clusters identified to date. We present, for the first time, the derived number density of the galaxy clusters to a flux limit of $1 \cdot 10^{-14}$ erg cm$^{-2}$ s$^{-1}$ (0.5–2.0 keV). This extends the $\log N$–$\log S$ of previous cluster surveys by more than one decade in flux. Results are compared to theoretical predictions for cluster number counts.

*Subject headings:* galaxies: clusters: general — X-rays: galaxies, general — techniques: image processing

---

[1] Dept. of Physics, Università di Roma *La Sapienza* (I.C.R.A.)
[2] Space Telescope Science Institute, Baltimore, MD, 21218
[3] European Southern Observatory, Garching, Germany

## 1. Introduction

Clusters of galaxies are powerful tools for studying the large scale structure and evolution of the universe. Current data provide conflicting constraints on cluster evolution as well as on possible cosmological models depending on whether cluster samples are optically or X–ray selected (see e.g. the discussion in Briel & Henry, 1993). These discrepancies could reflect the well known problems which characterize optical studies, especially at high redshifts, such as chance coincidences of unvirialized systems, biases towards increasingly rich systems, and biases towards systems with anomalously bright galaxies. Optically selected X–ray observed cluster samples (Castander et al. 1994, Bower et al. 1994) still suffer from the biases of the original selection process. By using X–ray surveys (Gioia et al. 1990, Piccinotti et al. 1982) which have the advantage of selecting physical objects, deep virialized potential wells in the case of clusters, instead of projected systems, these difficulties can be significantly alleviated. X–ray selection is less affected by projection effects, is done entirely by objective criteria and is subject to biases which are more easily quantifiable than those inherent in optical selection, particularly when an appropriate detection technique is used to identify extended X–ray sources and estimate their structural parameters.

In this spirit, we have started a serendipitous search for galaxy clusters in deep ($T > 20$ ksec) ROSAT–PSPC pointed observations at high galactic latitude. The survey is being carried out using a new source detection technique for X–ray images based on the Wavelet Transform (Rosati, 1995). This approach has the unique capability to detect faint X–ray sources with greater sensitivity over a wide range in sizes and surface brightnesses when compared with standard detection methods; thus probing a larger range of cluster parameters.

## 2. Observations and Data Analysis

The data are 15 ROSAT–PSPC pointed observations drawn largely from the public archive. Stellar targets or survey fields are selected with $T > 20$ Ksec and $|b| > 20°$. For AGN targets only cluster candidates more than $5'$ from the field center are retained.

We have developed a fully automatic procedure for a homogeneous data reduction of the PSPC observations. An image is extracted in the hard band [0.5–2.0 Kev] with a small binning factor ($8''$ pixel size) to adequately oversample the PSF. An energy–dependent exposure and vignetting correction map is applied. A Wavelet Detection Algorithm (WDA) is then used to localize and characterize sources.

The Wavelet Transform (WT) has been extensively discussed in the literature (Grossman et al. 1988, Bijaoui et al. 1991) and has been applied to a variety of astronomical problems (e.g, Slezak et al. 1993, Coupinot et al. 1992). The method developed for the analysis of the X–ray images is described elsewhere (Rosati et al. 1993, Rosati 1995). Briefly, it consists of an orthonormal decomposition of a signal into both space and scale through a convolution with a family of functions called wavelets. The continuous WT of an image I(x,y) with respect to the analyzing wavelet $\psi_a(x,y)$ is the 3–dimensional set: $W(x,y,a) = \frac{1}{a}I(x,y) \otimes \psi^*(\frac{x}{a}, \frac{y}{a})$, where $a$ is the wavelet scale. The wavelet coefficient $W(x,y,a)$ gives information about the signal at the location $(x,y)$ for the scale $a$. In the Fourier space, the WT is interpreted as a multi-scale filtering process. The signal is examined both in ordinary space, pixel by pixel, and in the frequency space, spatial scale by spatial scale. An inversion formula exists to recover the original image from its wavelet coefficients.

A uniform sampling of the scale $a$ in logarithmic scale, $a_i = (a_0)^i$, maximizes the orthogonality of the decomposition process. The wavelet used in this analysis is the difference of gaussians: $\psi_a(r) = \frac{2}{a^2}[e^{-\frac{r^2}{a^2}} - \frac{1}{2}e^{-\frac{r^2}{2a^2}}]$. This multi-scale analysis leads to an estimate of the morphological parameters (extent and net counts) of the detected sources by fitting the wavelet coefficients with the WT of the model assumed for the sources. This fitting procedure is repeated around each local maximum and the sources lying above the 99.95% confidence level (roughly $4\sigma$) are retained. The final parameter estimate is obtained by weighting the structural parameters deduced from *all* the analyzed scales $a_i$. The assumed source model is a gaussian isotropic profile which provides a good approximation of the PSF of the PSPC within an off–axis angle ($\theta$) of $15'$, hence only sources with $\theta < 15'$ are retained. An anisotropic wavelet analysis (giving ellipticity and orientation parameters) is also possible and is used to provide a better evaluation of fluxes and angular sizes of the cluster candidates.

There are many advantages of the Wavelet Transform technique which make it particularly suitable for this project: i) the method is equally rigourous for sources of different size; ii) no artificial parameters (such as detection/background box size) are needed; iii) morphological parameters can be estimated without knowledge of the background; iv) background gradients do not substantially affect detection efficiency and parameter estimation; v) sources are resolved down to the resolution limit and no blending is introduced; and vi) information on significant substructure within complex objects is preserved.

In order to check the validity of our data analysis procedure, we show in fig.1 the $\log N$–$\log S$ of all the 601 point-like sources in the sample and note the excellent agreement with the results from a previous ROSAT deep survey (Hasinger et al. 1993, (H93)). It should be stressed that our survey reaches a flux limit of $3 \cdot 10^{-15}$ erg cm$^{-2}$ s$^{-1}$ which is comparable to that of the H93 survey, while using shallower fields. In addition, no corrections for biases of the source detection procedure were required down to this flux limit (we estimated these to be $\lesssim 10\%$ effect using simulations). This demonstrates the greater ability of the WDA over conventional detection techniques in characterizing sources in the low counts and high source density regime.

For the conversion between count rate and flux, a power law spectrum with energy index 1 for point-like sources and a thermal spectrum with $T = 6$ keV for extended sources have been assumed along with the galactic HI absorption appropriate for each field.

The sky coverage has been computed as follows. For each field, the minimum detectable count rate in each radial bin of $1'$ width is calculated by integrating the average background over the average 95% power radius of the fitted profile and using the adopted detection threshold. For extended sources, the sensitivity depends on the intrinsic extent as well. The limiting sensitivities and respective solid angles of these annuli for each of the 15 fields are then summed to obtained the cumulative survey area as a function of limiting flux and source extent (see fig.2). Above a flux of $\sim 2 \cdot 10^{-14}$ erg cm$^{-2}$ s$^{-1}$, the entire FOV (2.95 deg$^2$) of the 15 pointings contributes to the survey area for point-like sources. Complete sky coverage is reached at higher fluxes for sources of increasing extent.

## 3. Cluster sample selection

Discrimination of extended emission, i.e. sources with characteristic size larger than the PSF, is more efficient in the central area of the detector where the PSF is sharper. In order to minimize the effects of confusion, in addition to retaining only extended sources with $\theta \leq 15'$, we only used pointings shorter than 80,000 seconds.

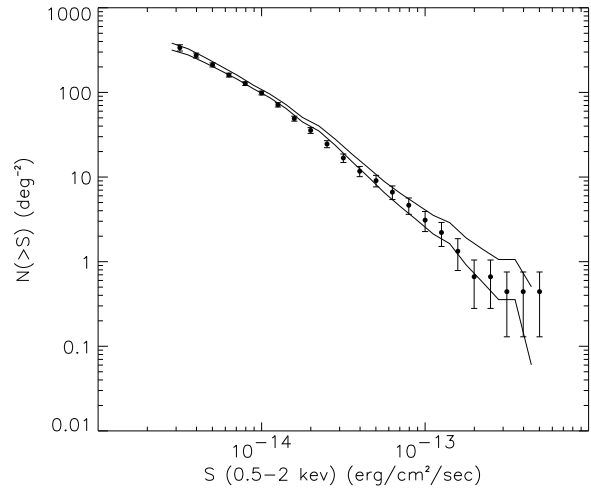

Fig. 1.— Cumulative number counts for all the X–ray sources of the sample (721 sources with $\theta \leq 18'$). Error bars ($1\sigma$) are from this survey, solid line is the $1\sigma$ band from H93 data (adapted from Comastri et al. 1995).

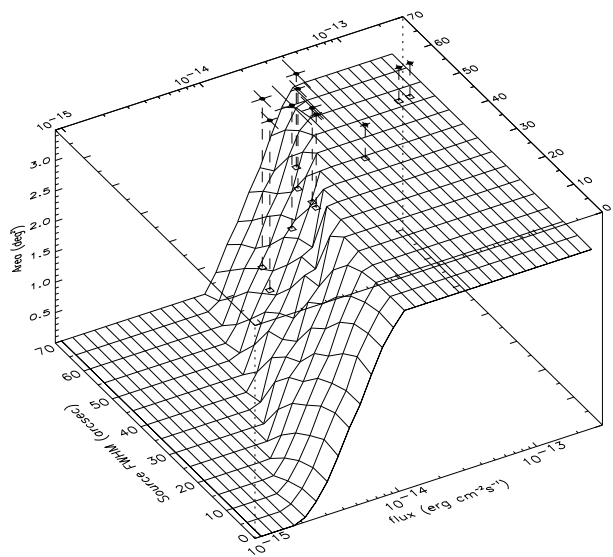

Fig. 2.— Sky coverage as function of flux limit and intrinsic source extent computed considering the central $15'$ of each field. Measured fluxes and intrinsic FWHMs of the 10 clusters in the sample are plotted on the top.

The FWHMs of the detected sources are compared with the best fit PSF as a function of the off-axis angle, as deduced from the whole set of evaluated sources. In order to identify possible systematic effects in estimated fluxes and sizes and to assess a significance for the sources to be extended, each field is simulated 10 times. This is done by replacing all the detected sources in the real fields with simulated sources drawn from a gaussian spatial distribution derived from the detected source parameters. The WDA is then re-run on the simulated images and the distribution of the FWHM residuals of the PSF best fit model is compared with the one observed in the real fields (see fig.3). The latter is a convolution of statistical uncertainties in the parameter estimate with the FWHM intrinsic spread due to different source spectra and $N_H$ absorption. From the distribution of FWHM residuals, one can derive a 99.9% c.l. for a source to be extended. Given the total number of detected sources, one can thus expect at most one source to be erroneously classified as extended as a result of random fluctuations. The final selection criteria for the cluster candidate sample are as follows: 1) Source FWHM > PSF FWHM at 99.9% c.l.; 2) $f_x \geq 1 \cdot 10^{-14}$ erg cm$^{-2}$ s$^{-1}$ = $f_{min}$.

X-ray emission on scales larger than $\sim 3'$ was not considered and hence corrections for background fluctuations (e.g. Snowden et al. 1994) were not required. However, there is very little sensitivity for detection of these very extended sources at these flux levels as can be seen from fig.2.

The flux cut $f_{min}$ allows us to maximize the completness of the flux limited sample as well as minimize the contamination from confusion effect. Using simulations with the observed $\log N$–$\log S$ down to fluxes of $10^{-15}$ erg cm$^{-2}$ s$^{-1}$, we estimate that at most 0.5 sources in our sample could be due to unresolved emission of two blended sources below $f_{min}$. This low level of contamination is a result of the WDA performing *adaptive aperture photometry*, hence the confusion effect only becomes significant at fluxes below $\sim 5 \cdot 10^{-15}$ erg cm$^{-2}$ s$^{-1}$. The same feature allows the flux to be integrated over the largest radius consistent with the background (which is very low for the case of the ROSAT–PSPC). For the clusters with measured redshift we have verified this to be at least 1 Mpc or approximately 4-5 core radii; thus limiting the lost flux to 10-20%, well within the flux errors. For the remaining clusters the R magnitudes of the brightest galaxies indicate a redshift interval between 0.2 and 1, thus providing again a physical radius of $\sim 1$ Mpc.

The selection criteria yield a sample of 10 cluster candidates in 15 processed fields. The brightest candidate has a flux of $1.8 \cdot 10^{-13}$ erg cm$^{-2}$ s$^{-1}$ ($\sim$400 counts). Inspection of POSS–E plates show that the optical counterparts of these cluster candidates are too faint to be readily identified. Preliminary CCD imaging of 7 candidates has revealed that at least 6 show a visible enhancement in the projected galaxy density at the peak of the X-ray emission. In addition, two further candidates turned out to be groups detected previously in other surveys (one at $z = 0.34$ in the Lockman field (M.Schmidt, private comunication) and the other at $z = 0.13$ in the Pavo field (Griffiths et al. 1983) ) and a third is a serendipitous rediscovery of a cluster in the Couch et al. (1991) distant cluster sample (J1888.16CL at $z = 0.563$). There are three more cluster candidates visible in fig.3 which have measured fluxes $(0.7 \leq f_x < 1) \cdot 10^{-14}$ erg cm$^{-2}$ s$^{-1}$ of which two show a significant overdensity of galaxies around the X-ray peak. These preliminary optical follow-up observations indicate that our technique has a high success rate. We show in fig.5 [plate 1] two of the optical images and overlaid X-ray contours.

## 4. Results: Cluster Number Counts

With this sample, we have measured the surface density of clusters at fluxes about an order of magnitude fainter than the limits reached by the Einstein Medium Sensistivity Sensitivity Survey (EMSS) (Gioia et al. 1990, Henry et al. 1992 (HY92)). We have derived a surface density of $6.7^{+2.9}_{-2.0}$ clusters per square degree at a flux limit of $1 \cdot 10^{-14}$ erg cm$^{-2}$ sec$^{-1}$. The errors correspond to the 68% confidence level and are deduced from the convolution of the dominating poisson statistics of the counts, N(>S), with the error distribution in N(>S) produced by the uncertainties in the estimated fluxes and sizes as determined via simulations. In fig.4, we have plotted a compilation of observed number counts over 4 decades in flux. The number counts from the EMSS sample have been obtained by integrating the luminos-ity functions (XLF), $n(L) = K(z)L^{-\alpha(z)}$, reported in HY92 over the luminosity range $(10^{43.5} - 10^{45})$ erg s$^{-1}$ in the three published redshift bins for $z \geq 0.14$ and using the Piccinotti et al. (1982) XLF for $z < 0.14$, i.e. $\alpha = 2.15$, $K = 3.8 \cdot 10^{-7} Mpc^{-3} L_{44}^{-1}$ (converted to the Einstein band and in agreement with the EMSS XLF within the systematic errors as discussed in HY92) ($H_0 = 50$ km s$^{-1}$ Mpc$^{-1}$, $q_0 = 0.5$ have been adopted). The integration is performed for two possible core radii (0.2 and 0.3 Mpc) which defines the closed area in fig.4 once the fluxes are converted to the ROSAT hard band (assuming a $T = 6$ keV thermal spectrum). Point 2 is from an X-ray selected sample of clusters from the ROSAT All Sky Survey (RASS) in the northern sky whose optical identification is 95% complete (Burg et al. 1994a). For reference, the derived number count from the Piccinotti sample is also plotted (point 3).

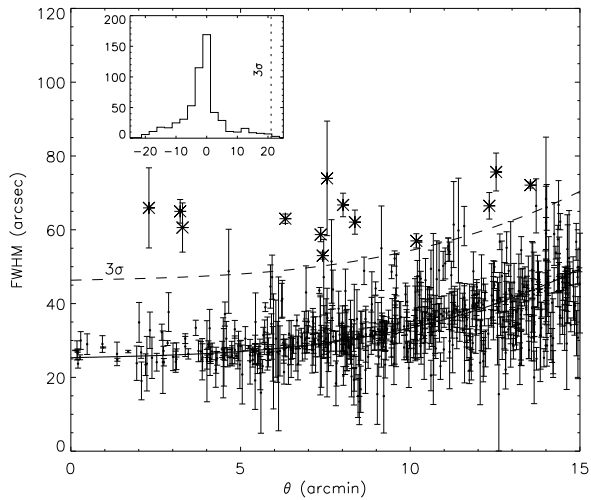

Fig. 3.— Estimated FWHM and 90% errors for all 601 sources vs off-axis angle. The stars indicate cluster candidates; the solid line is a polynomial fit to all data points clipped at $2\sigma$. The inset shows the distribution of FWHM residuals. The 3 sigma lines were determined from Monte-Carlo simulations.

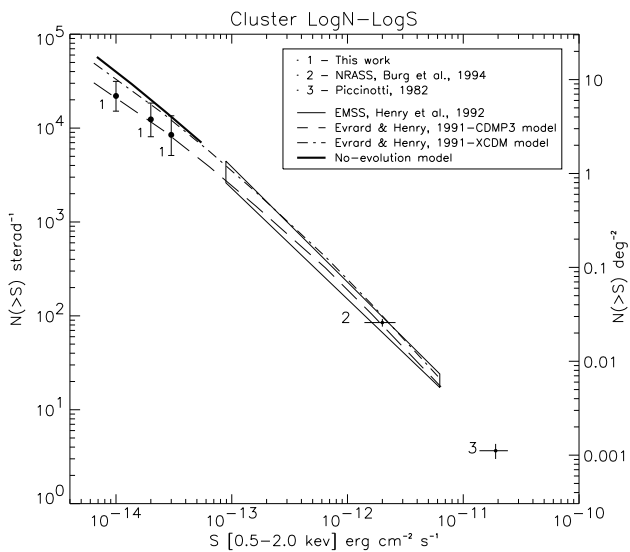

Fig. 4.— Observed and theoretical cluster cumulative number counts compiled from various sources.

We should note that the surface density of clusters at low flux levels in fig.4 should be considered a lower limit since the cut imposed on the extent excludes clusters from the sample which are unresolved by the PSPC. Assuming a standard King profile for the surface brightness of clusters with $\beta < 0.8$ and core radius $r_c = 250$ kpc, one can show that the 50% power radius exceeds $30''$ up to $z = 1$ in a critical universe with $H_0 \geq 50$ Km s$^{-1}$ Mpc$^{-1}$. Therefore, high redshift groups could possibly be missed in our sample. The presence of cooling flows with a $\Sigma$ excess of 10–30% in the center does not decrease the ability of the WDA to detect and classify sources as extended. However, extremely bright AGN in the center could result in a misclassification. The completeness function could also be affected if cluster parameters such as $r_c$ significantly change over the mass and redshift ranges probed.

Results from ROSAT medium sensitivity surveys, whose optical identifications are now $\sim 90\%$ complete, can be used to place limits on the possible incompleteness of our sample due to misclassification of unresolved sources. These surveys, which utilize detection algorithms more geared to the detection of point-like sources, are finding cluster number counts generally lower than our survey by a factor 5 at $f_x = 2 \cdot 10^{-14}$ erg cm$^{-2}$ s$^{-1}$ (Georgantopoulos et al. 1995). Hence, we expect the incompleteness of our sample due to unresolved clusters to be less than 20%.

## 5. Discussion

In fig.4, we also overplot theoretical predictions for cluster number counts according to two models by Evrard & Henry, 1991 (EH91). The two curves have been shifted in the [0.5–2.0]keV band and rescaled to allow for the different normalization used for the XLF in the lowest redshift shell by a factor $K_{EH91}/K_{z=0} = 7.2/3.8 = 1.9$. These curves represent a CDM model (spectral index of the primordial fluctuations, $n = -1$) with an $L \propto M^3(1+z)^{7/2}$ luminosity–mass relation (CDMP3) and a model with more power on large scales ($n = -2$) and $L \propto M^{11/6}(1+z)^{11/4}$ (XCDM). Both models incorporate the negative evolution of the XLF as observed in the EMSS sample (see also Cavaliere et al. 1993 for a physical interpretation of alternative models). Also shown in fig.4 is a curve representing a no evolution model which has been calculated by integrating the local ($z < 0.14$) XLF in the form of a power law down to luminosities of $2.5 \cdot 10^{43}$ erg$^{-1}$ s$^{-1}$ and up to $z = 1.5$ using $\alpha = 2.15$ and $K = 3.8 \cdot 10^{-7} Mpc^{-3} L_{44}^{-1}$ (see above). If a flattening of the XLF of X–ray selected samples is present, as has been observed in the XLF of optically selected samples (e.g. Burg et al. 1994b), then the same no evolution model would have a factor of two fewer counts at the survey flux limit and thus closer to the measured counts.

At present, small number statistics and the uncertainty in evaluating the completness of our sample prevent us from ruling out any of these models. More rigorous constraints will be possible once the redshift distribution and the volume density of our entire sample are known and compared with the results from complementary studies, such us the RASS, which probe the XLF on a wider range of luminosities and at lower redshifts. Our ultimate aim is to use our deep survey to study the faint end of the XLF at moderate–high redshifts thus addressing a crucial missing link in our current understanding of cluster evolution.

We are grateful to A.Ferguson, J.Huchra, B.McLean, I.Gioia, P.Henry and M.Schmidt for providing preliminary CCD observations. We are grateful to G.Hasinger for useful discussions on unpublished results from the RIXOS project. RB, RG and PR acknowledge partial support from NASA grants NAG5-1538, NAGW-2508, NAG8-794 and RDC acknowledges support from NAGW3288.